\documentstyle[aps,eqsecnum,epsfig]{revtex}
\begin{document}
\draft
\title{Self- generated disorder and  structural glass formation in
homopolymer globules}

\author{ V. G. Rostiashvili, G. Migliorini
and T. A. Vilgis$^{1,2}$}
\address{$^1$ Max Planck Institute for Polymer Research\\
  10 Ackermannweg, 55128 Mainz, Germany.\\
  $^2$ Laboratoire Europ\'een Associ\'e, Institute Charles Sadron\\
  6 rue Boussingault, 67083 Strasbourg Cedex, France.}

\date{\today}

\maketitle

\begin{abstract}
  
  We have investigated the interrelation between the spin glasses and the
  structural glasses. Spin glasses in this case are random magnets without
  reflection symmetry (e.g. $p$ - spin interaction spin glasses and Potts
  glasses) which contain quenched disorder, whereas the structural glasses are
  here exemplified by the homopolymeric globule, which can be viewed as a
  liquid of connected molecules on nano scales.  It is argued that the
  homopolymeric globule problem can be mapped onto a disorder field
  theoretical model whose effective Hamiltonian resembles the corresponding
  one for the spin glass model. In this sense the disorder in the globule is
  self - generated (in contrast to spin glasses) and can be related with
  competitive interactions (virial coefficients of different signs) and the
  chain connectivity. The work is aimed at giving a quantitative description
  of this analogy. We have investigated the phase diagram of the homopolymeric
  globule where the transition line from the liquid to glassy globule is
  treated in terms of the replica symmetry breaking paradigm. The
  configurational entropy temperature dependence is also discussed.

\end{abstract}


\section{Introduction}

The existing dynamical theory of the glassforming overcooled liquids and
polymers are mainly applicable to the relatively high temperature area. The
well-known mode - coupling theory (MCT) \cite{1} predicts a dynamical phase
transition at the critical temperature $T_{c}$ where the overcooled liquid
becomes nonergodic and stays in this state upon the further cooling. It is
significant that according to MCT, the characteristic times of the $\alpha $ -
relaxation or the viscosity exhibit a power low divergence: $\tau _{\alpha
  }(T)\propto \eta (T)\propto \left| T-T_{\rm c}\right| ^{-\gamma }$ at a
critical temperature $T=T_{\rm c}$.  It has been realized now that such types
of singularities are generic for a broad class of mean - field (MF) models and
its appearance is an indication for the role of the \textit{ activation
  processes} which are neglected in MF - models \cite{2}. This means
physically that the topography of the free energy landscape in the space of
the coarse grained variables should be taken into consideration.  At low
temperatures the free energy landscape determines more or less the dynamics of
the system.  The size of the free energy barriers between the metastable
states, however, determinates the rate of any activation processes at the low
temperature regime $T<T_{\rm c}$.

At the so-called Kauzmann temperature $T_{\rm K}$, which is usually $T_{\rm
  K}<T_{\rm c}$ a genuine thermodynamic glass transition or so called Gibbs -
DiMarzio transition is expected to take place \cite{3}.  At the Gibbs -
DiMarzio (or Kauzmann) temperature $T_{\rm K}$, the system is preferably in a
thermodynamic unique configuration since the height of the free energy
barriers grow with the system size.

The earliest analytical approach to the problem which captured these two
aforementioned aspects has been worked out in several papers by Kirkpatrick,
Thirumalai and Wolynes in the late 80's \cite{4}. In these papers which study
the phenomena by making use of the $p$ - spin interaction spin glasses (at $p
> 2$) and Potts glasses with more then four components, the following
conclusions have been drawn:

\begin{itemize}
\item The predicted phase transition temperature $T_{\rm A}$ obtained by the
  dynamical theory (actually equivalent to $T_{\rm c}$ of the MCT based on the
  Langevin dynamics in the mean field limit) is higher than the Kauzmann
  $T_{\rm K}$ obtained by the static theory (or by the ordinary replica
  theory).
  
\item As the temperature decreases (starting from the paramagnetic phase) the
  metastable minima appear first exactly at $T_{\rm A}$.
  
\item In the intermediate temperature regime $T_{\rm K} < T < T_{\rm A}$ many
  metastable states are separated by high barriers. Therefore these states may
  have a long lifetime. The activated transport is the typical process in this
  temperature range.
  
\item The overlap order parameter within the one-step replica symmetry
  breaking (1-RSB) scenario undergoes a discontinuous jump at $T_{\rm K}$. This
  reminds to a first order phase transition even so thermodynamically (e.g.,
  in the specific heat) the transition seems to be of second order. The
  authors called this class of phase transitions \textit{random first order
    phase transition}. Crisanti and Sommers confirmed essentially the same
  type of behavior for the $p$ - spin spherical model \cite{5}, which is
  different from the Sherrington - Kirkpatrick model, and is shared by the
  other spin - glass models without reflection symmetry \cite{6}.
\end{itemize}

Indeed it has been argued often, that many connections exist between the
behavior of the \textit{structural glasses} and spin glasses without
reflection symmetry \cite{2,4,6}. Most of these properties mentioned above for
the spin glasses can be found in the context of structural glasses too.
However, the important difference between the spin and structural glasses is
that the spin glasses models contain a \textit{quenched disorder } already in
the Hamiltonian. In structural glasses the Hamiltonian is a regular function
of the particles coordinates and does not contain disorder.  Nevertheless the
free energy in both systems resemble each other and posses similar properties.
Thus, and disorder is in a sense \textit{self - generated} and develops itself
during the cooling (or glass transition) process.

The properties and possibilities of self - generated (or self -induced)
disorder have been already discussed in the framework of rather special spin -
models \cite{2,7}. These models explicitly involve pseudo - random numbers,
i.e., the spin -spin couplings, although deterministic, oscillate very rapid
and can be considered as a (pseudo) - random variables. These models, however,
provide some spin analogies of structural glasses but still are quite
different from structural glasses \cite{2}.

Furthermore, the ideas and the methods of the mean - field spin - glasses
\cite{9,9',10} have been successfully used to study the freezing states in a
\textit{heteropolymer globule} \cite{Orl,Shakh1,Shakh2,Gros1}. The
corresponding model considers the conformational behavior of a polymer chain
with randomly quenched interactions between monomers. These random copolymers
posses always a collapsed (or globular state) depending on temperature and
strength of the randomness.  It was successfully shown, that the disorder
results in a globular and eventually in a glassy globular state.  The
corresponding freezing is characterized by a transition between two phases:
one phase is characterized by many accessible configurations, while the other
is dominated by only a few of them,  i.e., an
eventually frozen state.  In this context the heteropolymer freezing usually
serves as a simple ``toy model'' for the protein folding phenomenon
\cite{Gros1}. This relationship between the freezing and folding phenomena is
caused mainly by the fact that in both cases only a single conformation (which
is in the context of protein folding called the native state) dominates.

In the following we are going to investigate a similar problem in homopolymer
globules in poor solvent. It is well known that homopolymers in poor solvent
form globules due to an attractive second virial coefficient which are then
stabilized by the repulsive third virial \cite{collaps}. The relevant
parameter for the globule formation is the second virial coefficient which is
measured by the relative distance from the so-called $\Theta$ - temperature,
where the homopolymer takes Gaussian conformations.  Close below $\Theta$ the
polymer collapses and the deeper the temperature the more dense becomes the
globule. The intuitive limit is then a dense liquid globule. We will show
nevertheless, that even at lower temperatures and beyond a certain density the
globule freezes and forms a glass.  Homopolymer globules show a \textit{\ 
  structural glass transition}. Thus they will form a ``nano glass''.

It is most interesting to note that in recent Monte - Carlo simulations of
homopolymer globules, Milchev and Binder \cite{Milch} have found a dramatic
decrease of the acceptance rate of the moves which in its turn suggests to the
growth of the characteristic times. They have seen also a pronounced density
fluctuations which appear in the center of the globule and spread over the
entire globule as a temperature is lowered. This results have been interpreted
in terms of a glassy type of transition. These investigations have been
restudied in papers by Kreitmeier et. al \cite{Kreit1,Kreit2,Kreit3} where a
similar dynamical behavior of the homopolymer globule has been verified once
more and was generalized for the cyclic deformation regime.  Moreover, a glass
transition in a homopolymer globule (for the chain length $N < 27$ on the
$3\times 3 \times 3$ cubic lattice) has been discussed by computer simulations
\cite{Gros}. These authors found that the conformation space of the model
consist from one huge valley and a big number of small ``chambers''. These
disjoint domains of the conformational space are responsible for the
ergodicity breaking and freezing transition. This might be important for the
final processes of folding dynamics in the protein molecules because the speed
and reliability of the folding depends on whether the native state belongs to
the same conformational space domain or not \cite{Gros1}.

The thermodynamic behavior of the homopolymer globules have been studied also
by discontinuous molecular dynamics simulation \cite{Karplus1,Karplus2}. By
making use a simple off-lattice model for chain length $N = 64$ the authors
showed the existence of the first - order liquid - solid - like transition.
The transition occurs at temperatures below the coil - globule transition
temperature and has been detected by the heat capacity peak as well as by the
Lindemann parameter jump. It is interesting that there is a qualitative
similarity between homopolymers and proteins where the transition from the
molten globule to the native state is also of the first order
\cite{Karplus3,Ptitsyn}. In ref.\cite{Paul} a new ``expanded ensemble'' Monte
Carlo algorithm has been introduced which helps to overcome the density
slowing down in the globule state. For the chain length up to $N = 512$ the
authors have seen the bimodal distribution in the number of contacts per
monomer. This is again a clear indication of the first - order liquid - to -
solid transition of the collapsed globule.

As already mentioned, most important is that the globule formed from ordinary
homopolymers does not contain any quenched disorder. If this system forms
glassy states the disorder has to be self - generated. Nevertheless, to form
glassy states certain frustrations are necessary. These may have its origin in
the interplay between attractive interaction (negative $2^{\mathrm{nd}}$
virial coefficient $v$), repulsion (positive $3^{\mathrm{rd}}$ virial
coefficient $w$) and the constraints imposed by the chain connectivity. These
frustrations might be already sufficient to provide the structural glass
transition from a \textit{liquid globule} to a \textit{glassy globule}.

The main purpose of the present paper is to investigate these problems in more
detail by analytic means. We will indeed show that the linear homopolymer in
the condensed globular phase undergoes a genuine structural glass transition
of the similar type as in the $p$ - spin - interaction spin - glass with $p
>2$ or Potts glasses models. We stress once more, that the homopolymer is a
pure system and does not contain quenched disorder. Nevertheless the $n$ -
component field theory formulation for linear polymers \cite{degennes}
provides already a natural and solid basis \cite{vilgis91}.  As we will show
below, the homopolymer globule is a generic system where frustrations rather
than quenched disorder trigger the glass formation.  As a starting point we
employ the field - theoretical description of the self - interacted polymeric
chain in a poor solvent \cite{Gennes,Kholod}.  So far, in ref.
\cite{Gennes,Kholod} it has been shown that the Laplace transformation (with
respect to the chain length $N$) of the polymeric correlation function can be
associated with the corresponding correlator within the $n \longrightarrow 0$
limit of the $n$ - component $ \psi_{a}^4, \psi_{a}^6$ - field theory (where
$a = 1, \dots n $). We argue instead that the same $n$ - component field
theory can be mapped onto some kind of \textit{disordered one - component
  model} where the imposed random field is a non - Gaussian (a color noise)
and its statistical moments are determinated only by virial coefficients $v, w,
\dots$, etc.  In this case the index $a$ acts as a \textit{replica index} for
spin glasses \cite{9,9'} or random magnetics \cite{10}. This mapping has been
suggested for spin glasses already three decades ago \cite{11,11'} but was
never actually used.  At this point we should add a general remark: Throughout
the paper we use the {\it large globule approximation}, i.e., we consider the
chain length $N$ to be very large. This is necessary to avoid additional
complications close to the surface of the globule. Intuitively it is
obvious, that monomers close to the globule surface experience a different
dynamic environment, such that different mechanisms of freezing processes take
place. We will come back to this point in a later publication.

\section{The homopolymer globule as a disordered system}
\subsection{The field theory representation}

In this section we will provide the basic formulations and the field theory
for polymers in poor solvent, i.e., with attractive interactions.  To do so,
we start from usual continuous description of a homopolymer chain of the
length $N$ in the poor solvent. Its Hamiltonian has the following form
\begin{eqnarray}
H[{\bf r}(s)] =
\frac{d}{2a^2}\int\limits_0^N\, ds \left(\frac{\partial {\bf
      r}(s)}{\partial s}\right)^2 - \frac{|v|}{2}\int\limits_0^N\, dsds' \delta
\left({\bf r}(s) - {\bf r}(s')\right)\nonumber\\ 
+ \frac{w}{3!} \int\limits_0^N\, dsds'ds''\delta
\left({\bf r}(s) - {\bf r}(s')\right)\delta
\left({\bf r}(s') - {\bf r}(s'')\right),\label{corr}
\end{eqnarray}
where we allowed already that the $2^{\rm nd}$ virial coefficient is negative,
$v < 0$ and the $3^{\rm rd}$ virial coefficient $w > 0$. Usually the second
virial coefficient is measured by its distance from the $\Theta$ temperature,
i.e., $|v| \simeq a^3 |T-\Theta|/\Theta = a^{3} \tau$. The third virial
coefficient is always of order of $w \sim {\cal O}(a^{6})$. The density of the
globule is easily estimated by $\rho \simeq |v|/w$, which provides a
naive limit of the temperature at $\rho a^{3} = 1$. Indeed at sufficient low
temperatures the globule is dense, and the entropy term (first term) becomes
less and less important. The remaining entropic fluctuations are only important
on length scales $\xi \simeq a/ \tau$. This observation will allow us later
the safe use of corresponding mean field theories.

The next step is \cite{Gennes,Kholod,Panyukov} (see also Appendix A) to employ
the the Laplace transformation of the polymeric correlation function $\Xi({\bf
  r}_1,{\bf r}_2 ;N)$
\begin{eqnarray}
\Xi({\bf r}_1,{\bf r}_2  ;\mu) = \int\limits_0^{\infty} \, dN \, \Xi({\bf
  r}_1,{\bf r}_2  ;N)\exp (- \mu N)
\label{Laplace1}
\end{eqnarray}
which can be associated with the $n \to 0$ limit of the $\psi_{a}^4, \;
\psi_{a}^6$ field theory as follows:
\begin{eqnarray}
\Xi({\bf r}_1,{\bf r}_2  ;\mu) = \lim_{n\to 0}\int \, \prod_{a=1}^n 
\, D \psi_{a}\psi_{1}({\bf r}_1) \psi_{1}({\bf r}_2)\exp \left\{- 
  H_{n}[{\vec\psi};\mu] \right\}, \label{corr1}
\end{eqnarray}
where $\vec\psi = \{\psi_1, \psi_2, \dots,\psi_n \}$ is a $n$ - component
field, $\mu$ , the chemical potential, which is conjugated to $N$ and the
replicated Hamiltonian is
\begin{eqnarray}   
H_{n}[{\vec\psi};\mu] &=& \frac{1}{2} \int\, d^{d}r \sum_{a =
  1}^{n} \psi_{a}(r)\left[ \mu - \frac{a^2}{2d} \nabla^2
\right]\psi_{a}(r) - \frac{|v|}{8} \int\, d^{d}r \left[\sum_{a =
  1}^{n} \psi_{a}^{2}(r)\right]^{2}\nonumber\\
&+&\frac{w}{3!8}\int\, d^{d}r\left[\sum_{\alpha =
  1}^{n} \psi_{a}^{2}(r)\right]^{3} + \dots  \quad,\label{hamilt}
\end{eqnarray}
As usual the vector field $\psi$ corresponds to the polymer density in the
usual manner, i.e., $\rho \propto \langle \psi_{1}^{2} \rangle$
The relationship between $\mu$ and $N$ has the form
\begin{eqnarray}
N = - \frac{\frac{\partial}{\partial \mu}\int d^{d}r_1
  d^{d}r_2\Xi({\bf r}_1,{\bf r}_2  ;\mu)}{\int d^{d}r_1
  d^{d}r_2\Xi({\bf r}_1,{\bf r}_2  ;\mu)} \quad.
\label{Length2}
\end{eqnarray}
The attractive interaction term of the order $\psi^{4}$ changes the behavior
of the field theory. Correlations of the self avoidance are no longer
important and we must take care on the balance between the attractive and
repulsive forces. Although we mentioned above the relative unimportance of the
connectivity term, we have to take track on it as well. The careful analysis
below shows that it provides at sufficient low second virial coefficients
significant contributions on small scales ($\xi \sim {\cal O}(a)$)
contributions only by connectivity. These are in part responsible for
frustration.

\subsection{Mapping onto a random model}

Now we are going to map this field theory onto a random system. It can be
shown (see Appendix B) that the free energy of the globule state $F_{\rm Gl}$
can be interpreted as the free - energy of a one - component {\it random
  model} with Hamiltonian
\begin{eqnarray}
{\cal H} \{\psi\} = \frac{1}{2}\int\, d^{d}r\left[ \mu \psi^{2}(r) +
  \frac{a^2}{2d} \left(\nabla\psi\right)^2 + t(r) \psi^{2} (r) \right] 
, \label{hamilt1}
\end{eqnarray}
where the random field $t(r)$ is non - Gaussian with its generating functional
of the form
\begin{eqnarray}
\Phi \{\rho (r)\} &\equiv& \overline{\exp \left\{-\int\, d^{d}r\, t(r)\rho
    (r) \right\}}\nonumber\\
&=& \exp\left\{ \frac{|v|}{8}\int\, d^{d}r\, \rho^{2}(r) -
  \frac{w}{3!8}\int\, d^{d}r \, \rho^{3}(r)\right\},\label{GF}
\end{eqnarray}
and where the bar means the averaging over $t(r)$. It is interesting to
underline that only for this combination of signs of the virial coefficients
($v < 0, w > 0$) the even central moments $\overline{t(r_1) t(r_2) \dots
  t(r_{2m})}$ (see eqs.(\ref{moment2Ap}) - (\ref{moment5Ap})) are positive,
as it should be for the real field $t({\bf r})$.  The aforementioned mapping
takes the form
\begin{eqnarray}
F_{\rm Gl} = - \lim_{n\to 0} \frac{1}{n} \log \overline{Z^n},\label{free}
\end{eqnarray}
where the replicated partition function is
\begin{eqnarray}
\overline{Z^n} = \int \, D{\vec \psi} \exp \{ - H_n \}\quad. \label{partfun}
\end{eqnarray}
In the present paper we are going to use this analogy between the homopolymer
model (which is a ``pure'' model, i.e. does not include a quenched disorder in
its Hamiltonian) and a random model (\ref{hamilt1}). 

\subsection{Legendre transformation}

As a next step we should go to the two - replica - variables, $Q_{a b}(r)$, or
Parisi overlaps \cite{9} for polymers \cite{vilgis91}. One can merely
implement it using the Legendre transformation of the interaction part of the
Hamiltonian (\ref{hamilt}) which can be represented in the form
\begin{eqnarray}
K[u_{ab}] \equiv \frac{|v|}{8}\int\,
d^{d}r\,\sum_{a,b=1}^{n}\,u_{ab}u_{ba} -
\frac{w}{3!8}\int\, d^{d}r \,
\sum_{a,b=1}^{n}\,u_{ab}u_{bc}
u_{ca}\nonumber\\
-\frac{z}{4!16}\int\,d^{d}r \,
\sum_{a,b=1}^{n}\,u_{ab}u_{bc}u_{cd}u_{da} + {\cal O}(u_{ab}^5), \label{pair}
\end{eqnarray}
where the pair field $u_{ab}(r) \equiv
\psi_{a}(r)\psi_{b}(r)$ and where we have also  kept the $4^{\rm
  th}$ virial coefficient $z$.
Let us introduce the integral transformation 
\begin{eqnarray}
\exp\left\{K[u_{ab}]\right\} &=&\nonumber\\
&=&\int \,\prod_{c,d}^{n} D Q_{cd}(r)
\exp\left\{W[Q_{ab}] + \int\,d^dr \sum_{a,b}^n
  Q_{ab}(r)u_{ab}(r)\right\},
\label{transf}
\end{eqnarray}
and find $W[Q_{ab}]$ in the form of a functional expansion. For this purpose
one should use the saddle point method in eq.(\ref{transf}) which can be
carried out in the same spirit as in references \cite{Ohta,Fredrick}. This
results in the Legendre transformation with respect to the extremum field
$\overline{Q}_{ab}$
\begin{eqnarray}
K[u_{ab}(r)] = W[\overline{Q}_{ab}(r)] + \int\,
d^{d}r\,\sum_{a,b=1}^{n}\,\overline{Q}_{ab}(r)u_{ab}(r).
\label{Legendre}
\end{eqnarray}
As a result \cite{Zinn} we get:
\begin{eqnarray}
\frac{\delta \, K}{\delta u_{ab}} = \overline{Q}_{ab}(r) \quad,
\label{Legendre1}
\end{eqnarray}
\begin{eqnarray}
\frac{\delta \, W}{\delta\overline{Q}_{cd}(r)} = - u_{cd}\quad.
\label{Legendre2}
\end{eqnarray}
By making use of the expansion (\ref{pair}) in eqs.(\ref{Legendre1}) and
(\ref{Legendre2}) one obtain for $W[Q_{ab}]$ the following expression:
\begin{eqnarray}
W[Q_{\alpha\beta}(r)] = - \frac{2}{|v|}\int\,
d^{d}r\,{\rm Tr}(Q^2) &-& \frac{4w}{3|v|^3}\int\,
d^{d}r\,{\rm Tr}(Q^3) \nonumber\\
&-& \frac{2}{|v|^4} \left(\frac{w^2}{|v|} + \frac{z}{3}\right)\int\,
d^{d}r\, {\rm Tr}(Q^4) + {\cal O}(Q^5). \label{W}
\end{eqnarray}

After the transformation (\ref{transf}) the replicated partition function
(\ref{partfun}) takes the following form
\begin{eqnarray}
\overline{Z^n} = \int \,\prod_{c,d}^{n} D
Q_{cd}(r) \exp\left\{ -\frac{1}{2}{\rm Tr} \log\left\{[\mu - 
  \frac{a^2}{2d}\nabla^2] \delta_{ab} - 2
  Q_{ab}(r)\right\}  + W[Q_{ab}(r)] \right\}, \label{main}
\end{eqnarray}

So far only mathematical identities have been used. Nevertheless the use of the
overlap variables $Q_{ab}$ allow to detect completely different correlations
as the classical ${\cal O}(n)$ field theory in the limit $n \to 0$ for self
avoiding walks. They will allow to probe for a more complicated phase space
and provide information on the presence of glassy type correlations in the
globules. With this in mind it appears instructive to express the
generalization of the polymeric correlation function (\ref{corr1})
\begin{eqnarray}
\Xi_{ab}({\bf r}_1,{\bf r}_2  ;\mu) = \left< \psi_a({\bf r}_1)
  \psi_b({\bf r}_2)\right> 
\label{PolymCorr}
\end{eqnarray}
in terms of overlaps $Q_{ab}(r)$. For this end we add the source-term
$\psi_a(r)h_a(r)$ in eq.(\ref{transf}). After substitution in (\ref{partfun})
and integration over $\vec{\psi}$ one gets
\begin{eqnarray}
\Xi_{ab}({\bf r},{\bf r}'  ;\mu) =\left<\left\{\left[\widehat{1}\cdot
      G_0^{-1} - 2\widehat{Q}\right]^{-1}\right\}_{ab}\right>({\bf
  r},{\bf r}') ,
\label{PolymCorr1}
\end{eqnarray}
where $G_0^{-1}= \mu - (a^2/2d)\nabla^2$. The corresponding
polymer correlator (\ref{corr1}) is nothing but the $\Xi_{11}({\bf
  r},{\bf r}' ;\mu)$ element of the matrix (\ref{PolymCorr1}).

The correlator $\Xi_{ab}({\bf r},{\bf r}'  ;\mu)$ measures the
probability to find chain configurations started at ${\bf
  r}$ in the replica $a$ provided that it ends  at ${\bf
  r}'$ in the replica $b$. From eq.(\ref{PolymCorr1}) this correlator also 
can be seen as a scattering amplitude of a free ``particle'' with the
Green function $G_0$ on the ``scatterers'' whose   density is 
described by $Q_{cd}$.

The present  representation (\ref{main}) is very promising and reminds on
the corresponding expressions for the spin glass models \cite{3,5,9,9'}. The
striking difference between these two cases is that in eq.(\ref{main}) only
the pure model parameters (Kuhn's segment length and virial coefficients) are
involved. As a result the representation can provide a good starting point for
the phenomenon of {\it self - generated disorder } which,as we believe, is
behind the structural glass formation.

\section{Mean field treatment}

\subsection{Landau-type expansion}

In order to simplify the mean field treatment of the integral
(\ref{main}) let us expand the effective Hamiltonian in 
eq.(\ref{main}) to the $4^{\rm th}$ order. We then obtain by this
procedure
\begin{eqnarray}
\overline{Z^n} &=& \int \prod_{a,b=1}^n D
Q_{ab}(r)\exp\Biggl\{-\sum_{a,b=1}^n\int d^dr_1 d^dr_2
  \Gamma^{(2)}({\bf r}_1,{\bf r}_2)Q_{ab}({\bf r}_1)Q_{ba}({\bf
    r}_2)\nonumber\\
&-&\sum_{a,b,c=1}^n\int d^dr_1 d^dr_2 d^dr_3
  \Gamma^{(3)}({\bf r}_1,{\bf r}_2,{\bf r}_3)Q_{ab}({\bf r}_1)Q_{bc}({\bf
    r}_2)Q_{ca}({\bf r}_3)\nonumber\\
&-&\sum_{a,b,c,d=1}^n\int d^dr_1 d^dr_2 d^dr_3 d^dr_4
  \Gamma^{(4)}({\bf r}_1,{\bf r}_2,{\bf r}_3,{\bf r}_4)Q_{ab}({\bf r}_1)Q_{bc}({\bf
    r}_2)Q_{cd}({\bf r}_3)Q_{da}({\bf r}_4)\Biggr\} ,
\label{LandauExp}
\end{eqnarray}
where the coefficients
\begin{eqnarray}
\Gamma^{(2)}({\bf r}_1,{\bf r}_2) &=& \frac{2}{|v|}\delta({\bf r}_1 -
{\bf r}_2) - G_0({\bf r}_1 - {\bf r}_2)G_0({\bf r}_2 - {\bf r}_1)\nonumber\\
\Gamma^{(3)}({\bf r}_1,{\bf r}_2,{\bf r}_3) &=&
\frac{4}{3}\left[\frac{w}{|v|^3}\delta({\bf r}_1 - {\bf r}_2)\delta({\bf r}_2 - {\bf r}_3) - G_0({\bf r}_1 - {\bf r}_2)G_0({\bf r}_2 - {\bf r}_3)G_0({\bf r}_3 - {\bf r}_1)\right]\nonumber\\
\Gamma^{(4)}({\bf r}_1,{\bf r}_2,{\bf r}_3,{\bf r}_4) &=&
\frac{2}{|v|^4}\left(\frac{w^2}{|v|} + \frac{z}{3}\right)\delta({\bf
  r}_1 - {\bf r}_2)\delta({\bf r}_2 - {\bf r}_3)\delta({\bf r}_3 -
{\bf r}_4) \nonumber\\
&-& G_0({\bf r}_1 - {\bf r}_2)G_0({\bf r}_2 - {\bf
  r}_3)G_0({\bf r}_3 - {\bf r}_4)G_0({\bf r}_4 - {\bf r}_1).
\label{Coefficients}
\end{eqnarray}
As often in
the MF-theories the order parameter $Q_{ab}$ does not depend on the spatial 
coordinate  ${\bf r}$.
As it is  custom in the  MF-theory of the  spin glass models
\cite{9,9',10}, we decompose the Parisi matrix $Q_{ab}$ in the following
form
\begin{eqnarray}
Q_{ab} = (q - f)\delta_{ab} + f + \Delta_{ab}.
\label{Decompose}
\end{eqnarray}
In eq.(\ref{Decompose}) the symmetric part $R_{ab} = (q - f)\delta_{ab} + f$, 
with the diagonal $q$ and off-diagonal $f$ elements, describes the replica
symmetrical (RS) solution \cite{9,9',10}. The matrix $\Delta_{ab}$ equal zero
for $a=b$ and is responsible for the replica symmetry breaking
(RSB)\cite{9,9',10}. The use of the decomposition (\ref{Decompose}) in
eq.(\ref{LandauExp}) allows in the MF- approximation to separate the total
free energy into RS - and RSB - parts. During the calculation of the 
traces in
eq.(\ref{LandauExp}) it is convenient to use Parisi's representation of
$\Delta_{ab}$ by a function $\Delta(x)$, where $0\le x \le 1$. Then for the
free energy is found to be
\begin{eqnarray}
\lim_{n \to 0} \frac{1}{n V} \, F\{Q_{ab}\} = f_{\rm RS}\{q, f\}
  +f_{\rm RSB}\{q,f; \Delta(x)\},
\label{Decompose1}
\end{eqnarray}
where RS-free energy
\begin{eqnarray}
f_{\rm RS}\{q, f\} = A (q^2 - f^2) + B (q^3 -
3 q f^2 + 2 f^3) 
+C (q^4 - 6 q^2 f^2 + 8 
  q f^3 - 3 f^4) 
\label{RS}
\end{eqnarray}
and the RSB -free energy
\begin{eqnarray}
f_{\rm RSB}\{q,f; \Delta(x)\} = -{\rm w}_1 \int_{0}^1 dx
\Delta^2(x) - {\rm w}_2 \left[\int_{0}^1 dx \Delta(x) \right]^2 \nonumber\\
-{\rm
  w}_3 \int_{0}^1 dx \left[x \Delta^3(x) + 3 \Delta(x) \int_{0}^x dy
  \Delta^2(y)\right] + {\rm w}_4 \left[\int_{0}^1 dx
  \Delta(x)\right]^3\nonumber\\
+{\rm w}_5 \Biggl\{4\int_{0}^1 dx\Delta(x) \int_{0}^x dy y \Delta^3(y) 
+ \int_{0}^1 dx x^2 \Delta^4(x)\nonumber\\
 +2 \int_{0}^1 dx \Delta^2(x)\biggl\{
\left[\int_{0}^1 dy\Delta(y)\right]^2 + 2 \left[\int_{x}^1
  dy\Delta(y)\right]^2 - 2\int_{0}^x dy\Delta(y) \int_{y}^1 dz\Delta(z)\biggr\}\Biggr\} 
\label{RSB}
\end{eqnarray}
The coefficients in eqs.(\ref{RS}) and (\ref{RSB}) have the following
form (see Appendix C for the details)
\begin{eqnarray}
A &=& \frac{2}{|v|} - \left(\frac{d}{2\pi
    a^2}\right)^{d/2}\frac{\Gamma\left(\frac{4-d}{2}\right)}{\mu^{\frac{4-d}{2}}}\label{coeff1}\\
B &=& \frac{4}{3}\left[\frac{w}{|v|^3 } - \left(\frac{d}{2\pi
    a^2}\right)^{d/2}\frac{\Gamma\left(\frac{6-d}{2}\right)}{2\mu^{\frac{6-d}{2}}}\right]\label{coeff2}\\
C &=& \frac{2}{|v|^4}\left(\frac{w^2}{|v|} + \frac{z}{3}\right) - \left(\frac{d}{2\pi
    a^2}\right)^{d/2}\frac{\Gamma\left(\frac{8-d}{2}\right)}{3\mu^{\frac{8-d}{2}}} \label{coeff3}
\end{eqnarray}

\begin{eqnarray}
{\rm w}_1 &=& A + 3 B (q - f) + 6 C (q - f)^2 \label{wcoeff1}\\
{\rm w}_2 &=&  -3 B f - 4 C f^4 \label{wcoeff2}\\ 
{\rm w}_3 &=& - B - 4 C (q - f) \label{wcoeff3}\\
{\rm w}_4 &=& - 4 C f \label{wcoeff4}\\
{\rm w}_5 &=& - C.\label{wcoeff5}
\end{eqnarray}
The minimization of $f_{\rm RS}$ leads to the RS-solution, $q_m$ and $f_m$, whereas the maximization (as it is the case for
the spin glasses \cite{9,9',10}) of $f_{\rm RSB}$ results in the
RSB-solution in terms of the overlap matrix $\Delta(x)$. The 
coefficients in the Landau - expansion of $f_{\rm RSB}$ depend also from
the RS-solution.

In order to take into account the spatial correlation in RS - sector (see
Sec.V B), it is convenient to assume that the variable $q$ and the
coefficients $A,B$ and $C$ are weakly ${\bf k}$- dependent. In this case the
Landau - expansion is more involved and is given in Appendix C.

\subsection{Density in terms of $Q_{ab}$}

The essential issue is to express the globule density $\rho$ in terms of the
order parameter $Q_{ab}$. This allows to detect any glassy features inside the
globule and to distinguish between the liquid and glassy phase. Moreover it
will show any unusual properties of the phase space.  As mentioned already in
the introductory remarks the density is determined by the $\psi$ - fields.
Especially in the MF-approximation the local monomer density is given in terms
of the grand canonical partition function as
\begin{eqnarray}
\rho({\bf r}) = \int\limits_{0}^{N} ds \frac{\int d^d r \Xi({\bf r};s) \Xi({\bf
    r}-{\bf r}';N - s)}{\int d^d r \Xi({\bf r};N)}, 
\label{density}
\end{eqnarray}
which should be supplemented by the normalization condition (mass
conservation), i.e.,
\begin{eqnarray}
\int d^d r \rho({\bf r}) = N .
\label{normalization}
\end{eqnarray}
By making use of eqs.(\ref{density}) , (\ref{normalization}) and after Laplace
transformation we find
\begin{eqnarray}
\int d^d r \int d^d r' \Xi({\bf r}';\mu) \Xi({\bf r} - {\bf r}';\mu) =
-\frac{\partial}{\partial \mu}\int d^d r \Xi({\bf r};\mu) \quad,
\label{Convolution}
\end{eqnarray}
where $\Xi({\bf r};\mu)$ is the Laplace transformation of $\Xi({\bf r};N)$ (
see eq.(\ref{Laplace1})).  Then the equation which determines the chain length
becomes
\begin{eqnarray}
N = - \frac{\partial}{\partial \mu}\log \left\{\int d^d r \Xi({\bf r};\mu)\right\}
\label{Length}
\end{eqnarray}
and takes the form
\begin{eqnarray}
N = \frac{\int d^d r \int d^d r' \Xi({\bf r}';\mu) \Xi({\bf r} - {\bf
    r}';\mu)}{\int d^d r \Xi({\bf r};N)} .
\label{Length1}
\end{eqnarray}
In the MF- approximation eq.(\ref{corr1}) reads
\begin{eqnarray}
\Xi({\bf r}) = \psi_1^{\rm mf}({\bf r})\psi_1^{\rm mf}(0) ,
\label{MFcorr}
\end{eqnarray}
where $\psi_1^{\rm mf}({\bf r})$ is the MF-solution for $\psi_1({\bf
  r})$. Combining eqs.(\ref{MFcorr}) with (\ref{Length1}) and
  (\ref{normalization}) leads to the expected result
\begin{eqnarray}
\rho ({\bf r}) = \left[\psi_1^{\rm mf}({\bf r})\right]^2 .
\label{density1}
\end{eqnarray}
In order to express finally the density in terms of $Q_{ab}$ we recall that
the pair field $u_{11}^{\rm mf} = \left[\psi_1^{\rm mf}({\bf r})\right]^2$ (
see Sec.IIC).  Combining this with eqs.(\ref{Legendre2}) , (\ref{W}) and
taking into account the decomposition (\ref{Decompose}) one obtain for the
density the following expansion
\begin{eqnarray}
\rho &=& \frac{4}{|v|} q + \frac{4 w}{|v|^3}q^2 +
\frac{8}{|v|^4}\left(\frac{w^2}{|v|} + \frac{z}{3}\right)q^3 + 
\frac{4 w}{|v|^3} \left[ - \int_0^1 dx
  \Delta^2(x)\right]
+ \frac{8}{|v|^4}\left(\frac{w^2}{|v|} + \frac{z}{3}\right)\nonumber\\
&&\times\left[ -
  3q \int_0^1 dx\Delta^2(x) +  \int_0^1
dx\left(x\Delta^3(x) + 3\Delta(x)\int_0^x dy \Delta^2(y)\right)\right].
\label{density2}
\end{eqnarray}
In eq.(\ref{density2}) we have used Parisi's representation of
$\Delta_{ab}$ and taken into account that the off-diagonal element vanishes,
$f=0$ (see below). The ``singularity'' $|v|\to 0$ in eq.(\ref{density2}) is
spurious, as we will see in the next subsection  in the MF-approximation
 the value of $q$ becomes $|v|$ dependent itself, i.e., $q \propto |v|^2$
and $\Delta(x)\propto |v|^2$, so that $\rho \to 0$ at $v \to 0$ as it should
be close to the $\Theta$ - temperature.

\section{Theta point regime: coil-globule transition}

The question which must be resolved first is the ordinary coil globule
transition. In any case the present general approach should reproduce the
physical properties of the standard coil globule transition
\cite{degennes,GKh}. To do so we investigate the system close below to the
$\Theta$ - temperature $T < \Theta$ .  In ref. \cite{Gennes,Kholod} the
standard $O(n)$ - field theoretic formulation (see Sec.IIA) has been used in
order to treat this problem beyond scaling.  The method of pair fields, which
is a simpler version of the present formulation \cite{VilgHar} has been
developed and applied for the coil-globule transition. At that earlier paper
of one of the present authors the third virial coefficient has not been taken
into account, which corresponds to an expansion around the $\Theta$ point
regime.  Here, the Legendre transformation method from Sec.IIC allows easily
to take into account an arbitrary number of virial coefficients.

In the theta point region (i.e. at $T\le\Theta$) the globule conformations are
very close to Gaussian form so that for the chemical potential one can expect
the following scaling: $\mu = \mu_0/N$. In this case it is convenient to
rescale the virial coefficients $|v|, w$ and $z$ in the following way
\begin{eqnarray}
x = \frac{|v|}{a^d} \, N^{\frac{4-d}{2}} ,\label{Scaling1}\\
\nonumber\\
y =  \frac{w}{a^{2d}} \, N^{3-d} ,\label{Scaling2}\\
\nonumber\\
t =  \frac{z}{a^{3d}} \, N^{\frac{8-3d}{2}} ,\label{Scaling3}
\end{eqnarray}
which shows the upper critical dimensions of the different terms in the virial
expansion.  After that the scaling form of the coefficients (\ref{coeff1}) ,
(\ref{coeff2}) and (\ref{coeff3}) is given by
\begin{eqnarray}
A &=& \frac{N^\frac{4-d}{2}}{a^d}\left[\frac{2}{x} - \left(\frac{d}{2\pi
    }\right)^{d/2}\frac{\Gamma\left(\frac{4-d}{2}\right)}{\mu_0^{\frac{4-d}{2}}}\right]\quad,\label{Coeff1}\\
B &=& \frac{4\, N^\frac{6-d}{2}}{3\, a^{2d}} \left[\frac{y}{x^3 } - \left(\frac{d}{2\pi
    }\right)^{d/2}\frac{\Gamma\left(\frac{6-d}{2}\right)}{2\mu_0^{\frac{6-d}{2}}}\right]\quad,\label{Coeff2}\\
C &=&  \frac{N^\frac{8-d}{2}}{a^{3d}} \left[2\, \frac{y^2}{x^5} + \frac{2}{3}\, \frac{t}{x^4} - \left(\frac{d}{2\pi
    }\right)^{d/2}\frac{\Gamma\left(\frac{8-d}{2}\right)}{3\mu_0^{\frac{8-d}{2}}}\right] \quad.\label{Coeff3}
\end{eqnarray}
In the theta point regime $|v| = a^d (1 - T/\Theta) \to 0$ and $N >>
1$, so that at $d=3  \quad x \approx 1, \quad  y \approx 1$ and $t<<1$,
i.e. the forth virial coefficient becomes irrelevant.

In the present regime only RS-solution makes physical sense, since no other
solution than the onset of the liquid globule can be expected. Thus we
minimize RS-free energy (\ref{RS}) with respect to $q$ and $f$.  The resulting
solution reads
\begin{eqnarray}
f_m &=& 0 \label{Solution1}\\
q_m &=& \frac{-3 B + \sqrt{(3 B)^2 + 32 |A| C}}{8 C}, \label{Solution2}
\end{eqnarray}
so that the RS-free energy becomes
\begin{eqnarray}
f_{\rm RS}\{q\}  =  A\,q^2 + B\, q^3 + C \, 
  q^4 .\label{FreeEn} 
\end{eqnarray}
Let us consider the possible second order phase transition and impose the
following conditions: $A\le 0, B>0$ and $C > 0$. In the vicinity of the
transition point the coefficient $A$ becomes small $|A| << 1$, and the order
parameter takes the value
\begin{eqnarray}
 q_m \approx  \frac{2 \,|A|}{3\,B}.
\label{OrderPar}
\end{eqnarray}
As is seen from eqs. (\ref{Coeff1}) and (\ref{Coeff2}), in $d=3$ the order
parameter scales as $q_m \propto 1/N$. Thus, as it follows from
eq.(\ref{density2})  the density scales as
$\rho = 4 q_m/|v|$, which means physically $\rho = 1/N^{1/2}$
(note also $x\approx 1$ and $|v| \propto N^{-1/2}$). Therefore we reproduce 
the correct scaling for the density, which is found also from naive scaling.

We can also obtain the transition line in the $|v|$ - $w$ plane, which will be
the first step towards a more general phase diagram spanned by the virial
coefficients. This line is defined by the conditions $A=0, B > 0$ and $C > 0$,
which again for three dimensions $d=3$ yields
\begin{eqnarray}
\mu_0 &=&
\left(\frac{x}{2}\right)^2\left(\frac{3}{2\pi}\right)^3\left[\Gamma\left(\frac{1}{2}\right)\right]^2\quad,\label{condition1}\\
\nonumber\\
\frac{y}{x^3} &>& \left(\frac{3}{2\pi}\right)^{3/2}\frac{\Gamma
  \left(\frac{3}{2}\right)}{2 \, \mu_0^{3/2}} \quad,\label{condition2}\\
\nonumber\\
\frac{y^2}{x^5} &>& \left(\frac{3}{2\pi}\right)^{3/2}\frac{\Gamma
  \left(\frac{5}{2}\right)}{6 \, \mu_0^{3/2}} \quad.\label{condition3}
\end{eqnarray}
It is interesting that if eq.(\ref{condition1}) valid then conditions
(\ref{condition2}) and (\ref{condition3}) merge and convert into the globule
stability condition: $y > 16\pi^2/27$. We can eliminate $\mu_0$ in
eq.(\ref{condition1}) by combining eq.(\ref{Length}) with the polymer
correlation function
\begin{eqnarray}
\Xi({\bf k}; \mu) = \frac{1}{\frac{a^2}{6} k^2 + \mu - 2 q_m}.
\label{Corr3}
\end{eqnarray}
We recall that in eq. (\ref{Corr3}) $\mu =\mu_0/N$ and $q_m =
q_m^0/N$. The result of the combining reads 
\begin{eqnarray}
\mu_0 = 1 + 2 q_m^0 - 2 \, \frac{\partial}{\partial\mu_0} \,
q_m^0\quad .
\label{ChemPotential}
\end{eqnarray}
By making use of eqs.(\ref{OrderPar}) and (\ref{condition1}) in
eq.(\ref{ChemPotential}) for the transition line, one gets
\begin{eqnarray}
\left(\frac{x}{2}\right)^2\left(\frac{3}{2\pi}\right)^3 = 1 +
\left(\frac{4}{3}\right) \, \frac{\frac{16}{27} \, \pi^2}{y - \frac{16}{27} \,
  \pi^2} \quad ,
\label{TransitionLine}
\end{eqnarray}
where the globule stability condition $y > 16 \pi^2/27$ is implied.

For completeness we check for the possibility of a first order phase
transition. The necessary conditions for this are:$ A > 0 , B < 0$ and $C > 0$
(see eq.(\ref{FreeEn})).  It is simple to see by
eqs.(\ref{Coeff1}-\ref{Coeff3}) that these conditions are contradictory. This
means that within our MF-approach only the second order coil - globule phase
transition is possible which is in accordance with the well known result
\cite{GKh}. Therefore the present field theoretic formulation is able to
reproduce the standard coil globule transition as the replica symmetric
solution at conditions close to the $\Theta$ - temperature.

\section{Deeper in the globule state: liquid versus glassy regime}
\subsection{RSB solution in the globule}

In the following section we investigate the possibility of replica broken
solutions deeper in the globular state. The globule density from naive scaling
is given by $\rho a^{3}= a^{3}|v|/w= \tau$ and has natural limit at $\tau
=1$. Physically this limit corresponds to a dense globular state without any
solvent inside.  At temperatures below the coil - globule phase transition,
but still far above $\tau = 1$ the system is usually characterized by a
monomer - monomer correlation length $\xi < R_{\rm Gl} \propto N^{1/3}$. In
this case the chemical potential and the density are no longer $N$- dependent.
It can be seen from dimensional analysis and simple scaling arguments
\cite{degennes,Kholod,GKh} that the chemical potential scales as $\mu \propto
|v|^2/w \propto \tau^{2}$, the density $\rho_{\rm Gl}\propto |v|/w \propto
\tau$ and $\xi \propto 1/|v| \propto \tau^{-1}$.

In this regime  fluctuations can be still important unless $a \le \xi <<
R_{\rm Gl}$, where the MF solution, which we have discussed in Sec.III,
becomes valid. Here one can expect that because of competitive interactions
(negative second virial coefficient versus positive third virial one) and the
constraints imposed by the chain connectivity only a few conformations are
dominated. This could manifest itself as the glass transition long known for
spin - glass models \cite{9,9',10} and heteropolymers \cite{Gros1}. Formally
speaking, this transition shows itself as a nontrivial solution which
maximizes the RSB - free energy functional (\ref{RSB}).

Before doing this extremization let us find first the corresponding equation
for the chemical potential. In the RSB - case we should calculate the
$\Xi_{11}$ element from equation (\ref{PolymCorr1}) and substitute it into
(\ref{Length}). We use here the so called one step replica symmetry breaking
(1-RSB) scenario, which is generic for the glass transition in $p$ - spin model
\cite{4,5,6}, random - energy model \cite{Derr} and random heteropolymers
\cite{Gros1}. Then  Parisi's function $\Delta(x)$ is defined by
only two parameters
\begin{equation}
\Delta(x) = 
\left\{\begin{array}{r@{\quad,\quad}l}
0 &x < x_{0}\\ \sigma &x > x_{0}
\end{array}\right.
\label{RSB1}
\end{equation}
The interpretation of $\sigma$ and the break point $x_0$ is the following
\cite{9'}. Within the 1-RSB scenario all replicas are grouped into two
clusters so that $1 - x_0$ is the fraction of the replicas which belongs to
the overlap cluster with the overlap strength $\sigma$. The rest of the
replicas does not overlap.  After this simplification the inversion in
eq.(\ref{PolymCorr1}) can be done analytically (see eq.(AII7)  of
ref. \cite{MezPar}). After a straightforward
calculation one gets
\begin{eqnarray}
\Xi_{11}({\bf k};\mu) = \frac{1}{x_0 \left[\frac{a^2}{2d}k^2 + \mu -
    2q - 2(1 - x_0)\sigma\right]}
- \frac{1 - x_0}{x_0 \left[\frac{a^2}{2d}k^2 + \mu -    2q - 2\sigma\right]}\label{Inversion}
\end{eqnarray}
Insertion of eq.(\ref{Inversion}) in eq.(\ref{Length}) simply yields
\begin{eqnarray}
\mu - 2 q_m - 2(1 - x_0)\sigma_m = {\cal
  O}\left(\frac{1}{N}\right) ,
\label{ChemPot}
\end{eqnarray}
where $q_m$ and $\sigma_m$ are the solutions which extremize
the free energies (\ref{RS}) and (\ref{RSB}) correspondingly. With
1-RSB assumption (see eq.(\ref{RSB1})) the free energy (\ref{RSB}) becomes 
\begin{eqnarray}
f_{\rm RSB}(\sigma, x_0) = &-&{\rm w}_1 (1 - x_0)\sigma^2 + |{\rm w}_3|
(1 - x_0)(2 - x_0)\sigma^3 \nonumber\\
&-& |{\rm w}_5|(1 - x_0)(3 - 3x_0 + x_0^2)\sigma^4 .
\label{FreeEnerRSB}
\end{eqnarray}
It is convenient to represent the chemical potential in the form
\begin{eqnarray}
\mu = \frac{|v|^2}{s(|v|,w)} \quad ,
\label{mu}
\end{eqnarray}
where $s(|v|,w)$ is a function of $|v|$ and $w$. For convenience we switch to
dimensionless variables (by keeping for simplicity the same notations):
\begin{eqnarray}
\frac{v}{a^3} \to v \quad , \quad \frac{w}{a^6} \to w \quad , \quad
\frac{z}{a^9} \to z \quad ,\quad \frac{s}{a^6} \to s , \nonumber\\
\nonumber\\
A a^3 \to A \quad , \quad B a^3 \to  B \quad , \quad C a^3 \to C \quad.
\label{Dimension}
\end{eqnarray}
We also introduce the reduced (with a bar) values:
\begin{eqnarray}
{\bar A}&=& A |v| \quad , \quad  {\bar B} =  B |v|^3\quad , \quad  {\bar
  C} = C |v|^5 \quad , \quad {\bar q}_m = \frac{q_m}{|v|^2} \quad ,
\quad {\bar\sigma}_m = \frac{\sigma_m}{|v|^2},\nonumber\\ 
{\bar{\rm w}}_1 &=& {\rm w}_1|v| \quad , \quad{\bar{\rm w}}_3 = {\rm
  w}_3|v|^3\quad , \quad {\bar{\rm w}}_5 = {\rm w}_5|v|^5\quad .
\label{ABC}
\end{eqnarray}
After that, the equation (\ref{ChemPot})  for $s(|v|,w)$ (or for the chemical potential) 
takes the  compact form
\begin{eqnarray}
\frac{1}{s} = 2 {\bar q}_m + 2 (1 - x_0) {\bar\sigma}_m \quad, 
\label{equation_s}
\end{eqnarray}
where 
\begin{eqnarray}
{\bar q}_m  = \frac{-3{\bar B} + \sqrt{(3{\bar B})^2 + 32 {\bar |A|}
    {\bar C}}}{8 {\bar C}} \quad,
\label{Qu_m}
\end{eqnarray}
\begin{eqnarray}
{\bar\sigma}_m = \frac{3|{\bar{\rm w}}_3|(2 - x_0) + 
\sqrt{(3|{\bar{\rm w}}_3|(2-x_{0}))^2 -
    32{\bar{\rm w}}_1|{\bar{\rm w}}_5|(3-3x_0+x_0^2)}}{8|{\bar{\rm w}}_5|(3-3x_0+x_0^2)} 
\quad ,
\label{Sigm}
\end{eqnarray}
\begin{eqnarray}
{\bar{\rm w}}_1 &=& {\bar A} + 2 {\bar C} {\bar q}_m^2 \quad ,\label{w1}\\
{\bar{\rm w}}_3 &=& - {\bar B} - 4 {\bar C} {\bar q}_m \quad ,\label{w3}\\
{\bar{\rm w}}_5 &=& - {\bar C} \label{w5}
\end{eqnarray}
and the reduced coefficients reads
\begin{eqnarray}
{\bar A} &=& 2 - \left(\frac{3}{2\pi}\right)^{3/2}\Gamma
\left(\frac{1}{2}\right)s^{1/2} \quad,\label{A}\\
{\bar B} &=& \frac{4}{3}\left[ w - \frac{1}{2}\left(\frac{3}{2\pi}\right)^{3/2}\Gamma
\left(\frac{3}{2}\right)s^{3/2}\right] \quad,\label{B}\\
{\bar C} &=& 2 w^2 + \frac{2}{3} z |v| - \frac{1}{3}\left(\frac{3}{2\pi}\right)^{3/2}\Gamma\left(\frac{5}{2}\right)s^{5/2}\quad.\label{C}
\end{eqnarray}
Eqs.(\ref{Qu_m}) and (\ref{Sigm}) are the result of extremization
of eqs.(\ref{FreeEn}) and (\ref{FreeEnerRSB})
correspondingly. Throughout the remainder of the paper we will retain the
conditions : $A < 0 , B < 0 , C > 0 , {\rm w}_1 > 0 , {\rm w}_3 < 0$
and ${\rm w}_5 < 0$. This assures that the nontrivial solution
$\sigma_m$ shows up via  a first order phase transition, in a similar
manner as in $p$ - spin spin glasses \cite{4,5} or the random - energy 
model \cite{Derr}. On the coexistence line between liquid and glassy
phases  additionally $f_{\rm RSB}(\sigma_m , x_0) = 0$ and this leads
to the corresponding equation
\begin{eqnarray}
\frac{{\bar{\rm w}}_1 |{\bar{\rm w}}_5|}{|{\bar{\rm w}}_3|^2} = \frac{(2 - x_0)^2}{4 (3
  - 3x_0 + x_0^2)} \quad.
\label{Coexistence}
\end{eqnarray}
The equation for the reduced globule density, $c = \rho |v|$, can be
easily obtained from eq.(\ref{density2}) under the 1-RSB assumption
(\ref{RSB1}). The calculation yields
\begin{eqnarray}
c &&= 4 q_m + \frac{4w}{|v|^2} q_m^2 +
\frac{8}{|v|^3}\left(\frac{w^2}{|v|} + \frac{z}{3}\right)q_m^3 -
\frac{4w}{|v|^2} \sigma_m^2 (1 - x_0) +
\frac{8}{|v|^3}\left(\frac{w^2}{|v|} + \frac{z}{3}\right)\nonumber\\
&&\times\left[ - 3q_m \sigma_m^2 (1 - x_0) + \frac{{\sigma_m}^3}{2}(1 -
  x_0^2) + \frac{3}{2} {\sigma_m}^3 (1 - x_0)^2\right]
\label{ReducedDens}
\end{eqnarray}
Eqs.(\ref{equation_s}) - (\ref{C}) for the function $s(|v|,w)$ can be
solved numerically at given values of the  forth virial coefficient $z$ and 
break point $x_0$. After substitution of this solution $s(|v|,w)$ in
 eq.(\ref{Coexistence}) we arrive at the equation for the
coexistence line in the plane of $|v|$ and $w$. By changing $x_0$ one
can obtain a whole set of $x_0$ - isolines. We will give the
corresponding numerical solution in Sec.V C  but before we  analyze 
the validity of the MF - approximation which has been given above.

\subsection{Role of fluctuations in RS - sector}

As have been mentioned above, the MF-solution is valid when the
fluctuations are negligible. Generally speaking, this should be
required  for RS - and RSB - sectors of the replica space. In the
present paper it is not  our intention to consider fluctuations in the
RSB-sector, which is rather involved, and we leave it for a future
publication.

In the RS - sector spatial fluctuations are described by the
correlation function  (\ref{Correlator_q}) (see Appendix C). It is
easy to calculate from eq.(\ref{Correlator_q}) the radial distribution 
function $g(r) = 4\pi r^2 \langle \Delta q(r)\Delta q(0) \rangle$ at $d = 3$. Again
we turn to the reduced variables (\ref{mu}) - (\ref{ABC}). After the inverse
Fourier transformation of eq.(\ref{Correlator_q}), the radial
distribution function reads
\begin{eqnarray}
g(r) = \frac{r}{2
  \overline{X}}\exp\left\{-3|v|\sqrt{\frac{\overline{X}}{\overline{Y}}} \;r\right\}\quad,
\label{Radial}
\end{eqnarray}
where 
\begin{eqnarray}
\overline{X} = \overline{A} + 3{\bar q}_m \overline{B} + 6{\bar q}_m^2 \overline{C}
\label{Xbar}
\end{eqnarray}
and
\begin{eqnarray}
\overline{Y} = \frac{1}{24\pi}\left(\frac{3}{2}\right)^{3/2}
\left[s^{3/2} + 5{\bar q}_m s^{5/2} + \frac{63}{4}{\bar q}_m^2
  s^{7/2}\right]  \label{Ybar}
\end{eqnarray}
It is important to note here once more, that the quantities ${\bar X}$ and
${\bar Y}$ do not have any additional $|v|$ - dependence.

We estimate now the Ginzburg parameter, $\varepsilon_G$, (see e.g.
\cite{Klimontovich}) as the ratio of $g(r)$ at its maximum to $g_{m}^2={\bar
  q}_m^2|v|^{4}$.  Then for the Ginzburg criterion we have the following
equation
\begin{eqnarray}
|v|^2 = \frac{1}{6 e \varepsilon_G
  {\bar q}_m^2 \sqrt{\overline{X}(|v|,w)\overline{Y}(|v|,w)}} \quad,
\label{Ginzburg}
\end{eqnarray}
where $e$ is the Napier number. Eq.(\ref{Ginzburg}) for a reasonably small
$\varepsilon_G$ represents a line in the $(|v|, w)$ plane which separate
fluctuating and MF regimes. From now on we will call these two regimes as the
{\it liquid} globule and the {\it glassy} globule correspondingly.

From eq.(\ref{Radial}) the correlation length is given by
\begin{eqnarray}
\xi = \frac{1}{3|v|} \; \sqrt{\frac{\overline{X}}{\overline{Y}}} \propto
\tau^{-1} \quad,
\label{Xi}
\end{eqnarray}
which is qualitatively in line with the standard result
\cite{degennes,Kholod,GKh} and the correct scaling.

From the Gaussian approximation for the effective Hamiltonian in the RS case
(\ref{Fluctuations1}) it is easy to calculate the corresponding RS - free
energy, which takes the following form
\begin{eqnarray}
f_{\rm RS}\{{\bar q}_m\} = |v|^3 \left[\overline{A} {\bar q}_m^2 +
    \overline{B} {\bar q}_m^3 + \overline{C}{\bar q}_m^4\right] +
  \frac{T}{2}\int\frac{d^3\kappa}{(2\pi)^3} \log
  \left[|v|^2\overline{X} + \kappa^2 \overline{Y} \right] \quad. 
\label{FreeEnFluct}
\end{eqnarray}
The last integral in eq.(\ref{FreeEnFluct}) diverges at large ${\bf \kappa}$.
This ultraviolet divergence is of no significance as soon as the $|v|$ -
dependence is the only one we are interested in. For the second derivative of
the integral in eq. (\ref{FreeEnFluct}) with respect to $|v|^2$ one has
\begin{eqnarray}
I_{|v|^2}^{''} = -
\frac{T\overline{X}^2}{4\pi^2}\int\limits_{0}^{\infty}\;\frac{\kappa^2 
  d \kappa}{\left[|v|^2\overline{X} + \kappa^2\overline{Y}\right]^2}
= -
\frac{T}{16\pi}\left(\frac{\overline{X}}{\overline{Y}}\right)^{3/2} \frac{1}{|v|}\quad.
\end{eqnarray}
After that the expression for the RS - free energy takes the form
\begin{eqnarray}
f_{\rm RS} = |v|^3 \left[\overline{A} {\bar q}_m^2 +
    \overline{B} {\bar q}_m^3 + \overline{C}{\bar q}_m^4\right] -
  \frac{T}{12\pi} \left(\frac{\overline{X}}{\overline{Y}}\right)^{3/2} 
  |v|^3 \quad.
\label{FreeEnFluct1}
\end{eqnarray}
We ascribe this branch of the free energy to the liquid globule state
and will consider it in more details in the next subsection.

\subsection{Numerical calculations: chemical potential, phase diagram
  and configurational entropy}

Finally we are going to compute the phase diagram for the polymer globule. We
recall here that the first step carried out earlier in this paper, i.e., the
RS - solution corresponds to the classical coil globule transition. Here we
are now in the position to calculate from the 1-RSB free energy the transition
to the glassy state.  The numerical solution of eqs.(\ref{equation_s}) -
(\ref{C}) at $z=9$ and $x_0 = 0.88$ is shown in Fig.1 . As can be seen the
function $s(|v|,w)$ depends linearly from $w$ and almost does not depend from
$|v|$.  This is in agreement with well known result \cite{Kholod,GKh}: $\mu
\propto |v|^2/w$ (see eq.(\ref{mu})). We have also calculated $s(|v|,w)$ at
$x_0 = 0.90, x_0 = 0.92, x_0 = 0.95$ and have used these results as an input
in eq.(\ref{Coexistence}).  This eventually leads to $x_0$ - isolines in the
glassy globule phase which are plotted in Fig.2.

\begin{figure}[h]
  \begin{center}
    \includegraphics[scale=0.6,angle=0]{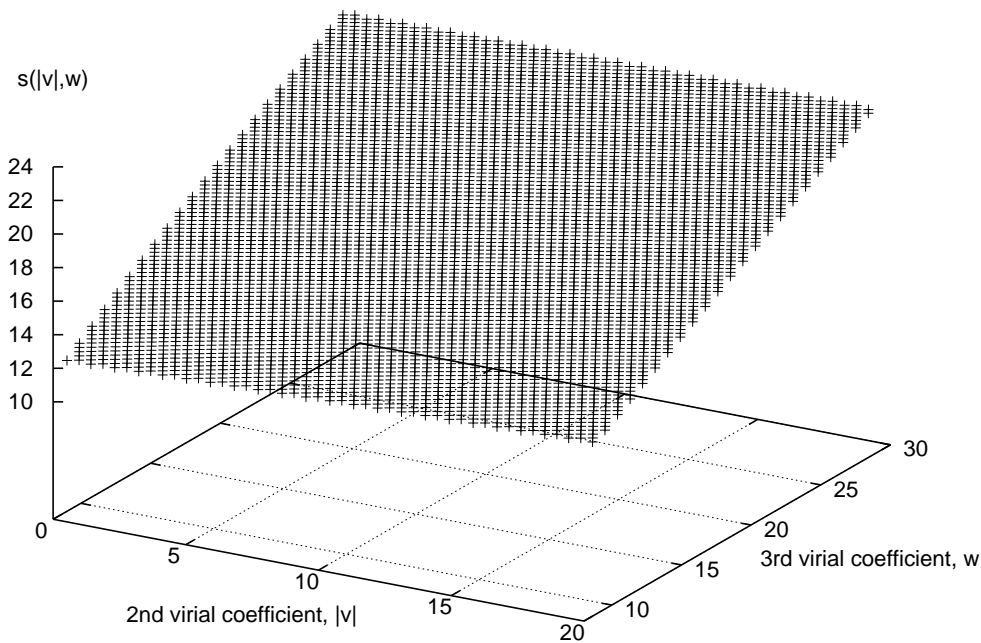} \vspace{5pt}
  \caption{The dependence of the parameter $s$ (see eq.(\ref{mu})) on the
    virial coefficients at $z = 9$ and $x_0 = 0.88$}
   \end{center}
\end{figure}

The line, which is associated with eq.(\ref{Ginzburg}) corresponds to the
Ginzburg criterion for fluctuations in RS-sector and separates the glassy
globule from the liquid one. Obviously, the position of this line depends from
the value of $\varepsilon_G << 1$ and should be better seen as a crossover
from the fluctuating regime to the mean field one. In Fig.2 this line is given
at $\varepsilon_G = 0.033$. We have not shown more $x_0$ - isolines
explicitly, but it is important to remind that by changing $x_0$ continuously
one can span up the whole phase digram from left to right. It is interesting
that the $x_0$ - isolines in Fig.2 are almost vertical. This means that in a
real experiment (upon changing the solvent quality $|v|$ by temperature)
always some particular value of $x_0$ is hit in the glassy phase and stays
with it as $|v|$ increases. We recall that $1 - x_0$ is the fraction of
replicas which overlap with the strength $\sigma_m = {\bar\sigma}_m |v|^2$.

\begin{figure}[h]
  \begin{center}
    \includegraphics[scale=0.6,angle=0]{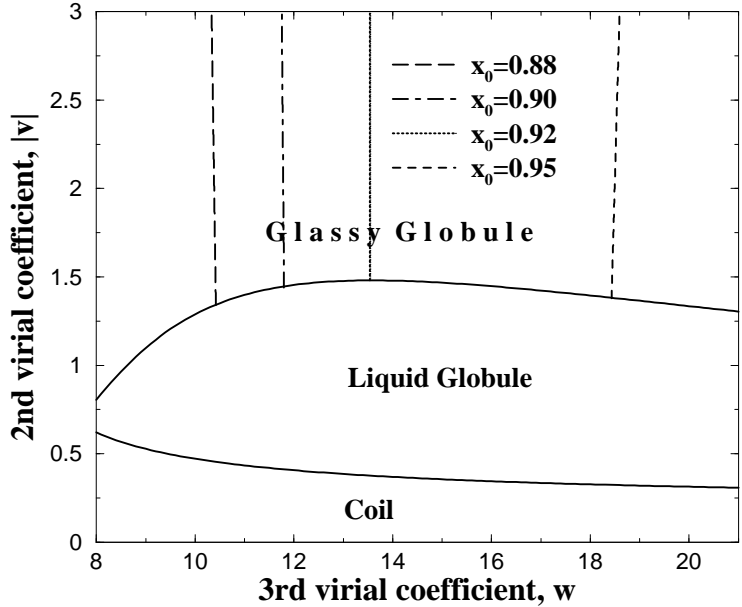} \vspace{5pt}
    \caption{The phase diagram of the polymer globule. The lower solid
      line corresponds to the coil - liquid globule second - order transition,
      whereas the upper solid line is associated with the first - order liquid
      - to - solid globule transition. Dotted (- dashed) $x_0$-isolines
      correspond to glassy state with different values of $x_0$. The other
      $x_0$-isolines which are not shown here have intermediate values of
      $x_0$.}
   \end{center}
\end{figure}

The fact that on the transition line the value of $x_{0}$ is less than one,
$x_0 < 1$ shows that the transition is thermodynamically of the first order.
This is contrary to  $p$ -spin spin glasses \cite{4,5,6} and random
heteropolymer \cite{Shakh1}. It is well known that in these cases transition
has no latent heat ( i.e.  thermodynamically of the second order) since $x_0 =
1$ at the transition point, whilst the order parameter $\sigma$ undergoes a
jump (i.e. displays the first - order transition). In our case the transition
is of the first order thermodynamically as well as with respect to the order
parameter. In Fig.2 we have shown also the line which corresponds to the coil
- globule second order transition (see eq.(\ref{TransitionLine})) at the chain
length $N = 250$. The critical value $|v|_{\rm cr}$ on this line is scaled as
$1/\sqrt{N}$.

Fig.3 shows the reduced density behavior (see eq.(\ref{ReducedDens}) at $x_0 =
0.88$ in the same as in Fig.2 intervals of $|v|$ and $w$. It can be seen that
the density on the transition line is fairly small to justify the use of the
virial expansion.

\begin{figure}[h]
  \begin{center}
    \includegraphics[scale=0.6,angle=0]{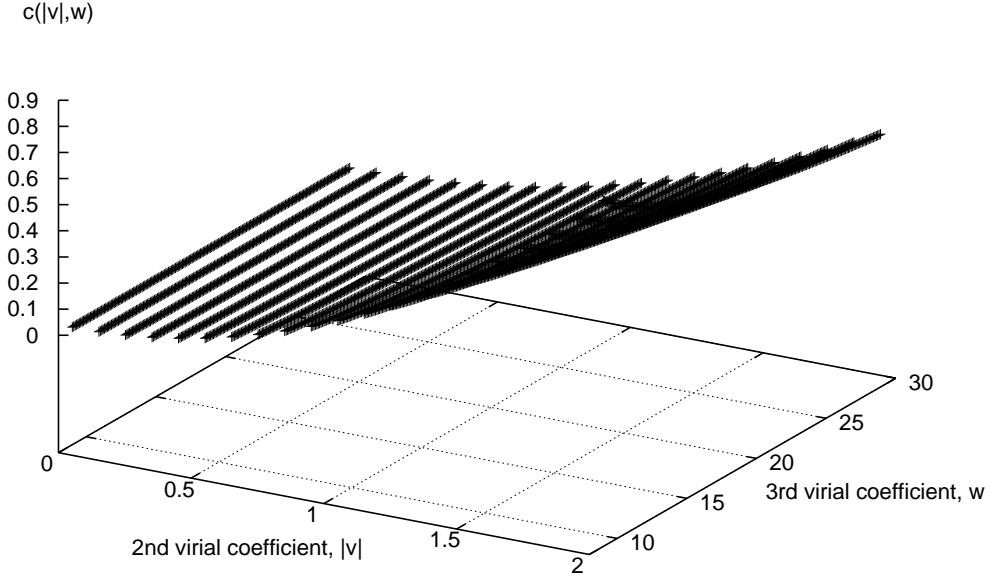} \vspace{5pt}
  \caption{The reduced density $c$ as the function of virial
    coefficients at $z = 9$ and $x_0$ = 0.88}
   \end{center}
\end{figure}

Let as now calculate the configurational entropy $S_{\rm conf}$ (or the
complexity) \cite{Palmer} which is usually of interest for glass forming
liquids. The configurational entropy in the liquid globule state can be
defined as the difference
\begin{eqnarray}
S_{\rm conf} = S_{\rm liquid} - S_{\rm valley}\quad,
\label{ConfEntrop}
\end{eqnarray}
where 
\begin{eqnarray}
 S_{\rm liquid} = - \frac{\partial f_{\rm RS}}{\partial T}
\label{LiquidEntrop}
\end{eqnarray}
and $S_{\rm valley}$ is the entropy (per particle) which corresponds to one
pure state or a valley in the free energy landscape. In order to estimate
$S_{\rm valley}$ let us recall from reference \cite{GrossMezard} that the
order parameter $\sigma$ describes the structure of space of valleys through
the probability, $P(\sigma)$, that two valleys picked at random, have an
overlap $\sigma$. For the 1-RSB - scenario this function has a rather simple
form: $P(\sigma) = x_0 \delta(\sigma) + (1 - x_0) \delta(\sigma - \sigma_m)$.
The dimension of the matrix $\Delta_{ab}$, which within 1-RSB is parameterized
by $n(x_0 - 1)/2$ parameters $\sigma$, becomes negative at $n \to 0$ and $0
\le x_0 \le 1$. Therefore the free energy $f_{\rm RSB}$ (see
eq.(\ref{FreeEnerRSB})) becomes negative, and  the factor $- (1 - x_0)$ in
eq.(\ref{FreeEnerRSB}) can be treated as  ``fraction of valleys'' with
overlap $\sigma_m$. In this case the intra-valley free energy can be obtain by
dividing out a common factor of $- (1 - x_0)$ in eq.(\ref{FreeEnerRSB}):
$f_{\rm valley} = - f_{\rm RSB}/(1 - x_0)$. Then we can estimate $S_{\rm
  valley}$ in a following way:
\begin{eqnarray}
S_{\rm  valley} = \frac{1}{1 - x_0} \; \frac{\partial f_{\rm
    RSB}}{\partial T} \quad.
\label{ValleyEntrop}
\end{eqnarray}

The configurational entropy in the glassy globule state is given by
the same way as in the $p$ -spin spin glasses \cite{4,GrossMezard}:
\begin{eqnarray}
S_{\rm conf} = \frac{1}{N}\;\left[\psi(1) - \psi(1 - x_0) \right]\quad,
\label{GlassEntrop}
\end{eqnarray}
where $\psi(x)$ is the digamma function. 

\begin{figure}[h]
  \begin{center}
    \includegraphics[scale=0.6,angle=0]{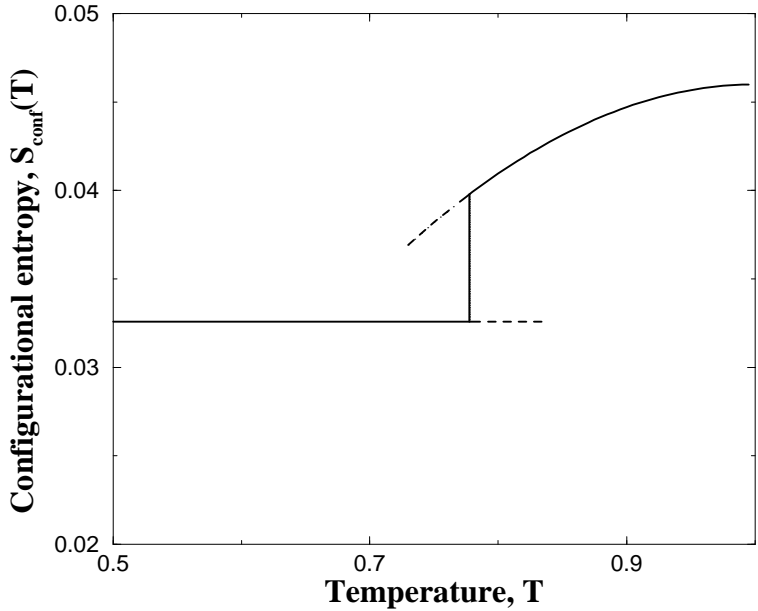} \vspace{5pt}
  \caption{The configurational entropy $S_{\rm conf}(T)$ as the
    function of temperature whilst the globule is cooling from the
    liquid to the glassy state}
   \end{center}
\end{figure}

By cooling the system along the $x_0$ - isoline at $x_0 = 0.88$, we have
calculated the configurational entropy for liquid (see eqs. (\ref{ConfEntrop})
- (\ref{ValleyEntrop})) and glassy ( eq.(\ref{GlassEntrop}) globule. Figure 4
gives the result of this calculations. As discussed above the transition is of
the first order, i.e., the $S_{\rm conf}$ undergoes a jump which is
qualitatively in line with Molecular dynamics \cite{Karplus1,Karplus2} and
Monte Carlo \cite{Paul} simulations. It must bear in mind that the
transition temperature in Fig.4 is also defined by the Ginzburg criterion for
fluctuations (see the intersection point of the upper solid line and the $x_0
= 0.88$ isoline in Fig.2), so that this transition can be treated as a
crossover from the fluctuating regime to mean field, where only few
states are dominated. This possibility is shown in Fig.4 by dotted line.  On
the other hand, this behavior is quite different from low molecular systems or
polymer melts where $S_{\rm conf}$ goes to zero continuously as soon as $T \to
T_{\rm K}$.

\section{Conclusion and outlook}

We have shown that the homopolymer globule problem can be formulated within
the $n \to 0$ limit of free $n$-component $\psi_a^4 , \psi_b^6$ - field theory
and then be mapped onto the disordered one - component model with a non -
Gaussian random noise. It is of interest, that the statistical moments of this
noise can be expressed only through the virial coefficients of the pure model,
the fact which gives grounds to discuss the self - generated disorder. On the
other side, the $n \to 0$ limit is responsible at the same time both for the
polymer conformations and for the nontrivial structure of the replica space.
Physically this means that units with competitive   interactions and stipulated
by the connectivity posses good pre-conditions for the glass formation.

The Legendre transformation from the pair fields $\psi_a(r)\psi_b(r)$ to the
Parisi overlap parameters $Q_{ab}(r)$ leads to the effective Hamiltonian which
is akin to corresponding expressions for the spin glass models. We have
investigated this resulting replicated model within mean field Landau - type
treatment.  First of all, the RS - solution of the corresponding equation is
associated with the conventional coil - liquid globule transition whereas the
RSB - solution in the deeper globule state is related with the glassy regime.
This mean field glassy globule phase can be only assured if the fluctuations
in RS - and RSB - sectors are small. In this paper we have only studied the
fluctuations in the RS - sector and have sketched the corresponding Ginzburg
criterion line on the phase diagram.  We have calculated the configurational
entropy for the liquid and glassy globules and shown that the transition
between them is a first order one, as also is in the MD and MC computer
simulations \cite{Karplus1,Karplus2,Paul}, whereas for the random
heteropolymer \cite{Shakh1} the freezing is thermodynamically a second order
transition. On the other side, the computer simulation \cite{Karplus3} and
experiment \cite{Ptitsyn} shows that the folding in proteins has a latent
heat, i.e. the random heteropolymer is a poor model of protein - folding
thermodynamics. It was shown \cite{Gros1,Ramanath} that in the heteropolymer
models where some sequences of monomers have especially law energy in their
native conformation the folding is thermodynamically a first - order
transition.

Nevertheless, our calculations shows that already for the homopolymer globule,
because of competitive interactions (virial coefficients of different signs)
and connectivity , the transition with a latent heat is possible. In this
respect the revisiting of the random heteropolymer problem within our field
theoretical approach (as oppose to the density functional method
\cite{Shakh1,Shakh2,Gros1}) would be very interesting. This could elucidate
the problem how the self - generated disorder interplay with quenched disorder
and eventually modify the freezing conditions.

The dynamical aspects of the thermodynamic transition discussed above
is a matter of crucial importance. The first and foremost question
which should be investigated is related with the formation of the
entropic droplets \cite{4,Takada} in the RSB - sector of the replica space. 

\section*{Acknowledgments}
V.G.R acknowledges greatfully the support by the Sonderforschungsbereich SFB
262 of the Deutsche Forschungsgemeinschaft.

\begin{appendix}
\section{Field-theoretical representation for a general
  self-interacting chain}
Let us represent the Hamiltonian of the self-interacting chain in the
following form
\begin{eqnarray}
H[{\bf r}(s)] &=& \frac{d}{2 a^2}\int\limits_0^N ds \left(\frac{\partial
    {\bf r}}{\partial s}\right)^2\nonumber\\ 
&+& \sum_{m=1}^{\infty}\frac{v_{m+1}}{(m+1)!}\int\limits_0^N ds_0 ds_1
\dots ds_m \prod\limits_{i=1}^m \delta \left({\bf r}(s_0) -  {\bf
    r}(s_i)\right) \quad,\label{Hamilt}
\end{eqnarray}
where $v_{m+1}$ denote the virial coefficients. It is convenient to
introduce the density 
\begin{eqnarray}
\rho ({\bf r}) = \int\limits_0^N  ds \delta \left({\bf r} -  {\bf
    r}(s)\right) , 
\end{eqnarray}
then the partition function of the polymer chain
\begin{eqnarray}
Z = \int D{\bf r}(s) \exp \left\{ - H[{\bf r}(s)]\right\}
\end{eqnarray}
takes the form 
\begin{eqnarray}
Z = \int D{\bf r}(s) D\rho ({\bf r})\delta \left[ \rho ({\bf r}) -
  \int\limits_0^N ds \delta \left({\bf r} -  {\bf
    r}(s)\right) \right] \nonumber\\
\times \exp \left\{-\frac{d}{2 a^2}\int\limits_0^N ds \left(\frac{\partial
    {\bf r}}{\partial s}\right)^2 -
\sum_{m=1}^{\infty}\frac{v_{m+1}}{(m+1)!}\int d^d r \rho^{m+1}({\bf r})\right\}.
\label{Partition}
\end{eqnarray}
By making use of  the integral representation for the $\delta$ - function
eq.(\ref{Partition}) can be recast in a form
\begin{eqnarray}
Z &=& \int D{\bf r}(s)D\phi ({\bf r}) D\rho ({\bf r})\exp \Biggl\{i \int d^d r \phi
  ({\bf r}) \rho ({\bf r})\nonumber\\
 &-& \sum_{m=1}^{\infty}\frac{v_{m+1}}{(m+1)!}\int d^d r \rho^{m+1}({\bf
   r})\Biggr\}\int d^d r d^d r' G\left({\bf r},{\bf r'};[\phi];N\right) ,
\label{Partition1}
\end{eqnarray}
where
\begin{eqnarray}
G\left({\bf r},{\bf r'};[\phi];N\right) = \int\limits_{{\bf r}(0)={\bf
    r}}^{{\bf r}(N)={\bf r'}}D{\bf r}(s)\exp\left\{-\frac{d}{2 a^2}\int\limits_0^N ds \left(\frac{\partial
    {\bf r}}{\partial s}\right)^2 - i\int\limits_0^N ds \phi ({\bf r}(s))\right\}\quad.
\label{Correlator}
\end{eqnarray}
The corresponding equation for $G$ reads
\begin{eqnarray}
\left[\frac{\partial}{\partial N} - \frac{a^2}{2 d}\nabla^2 + i\phi
  ({\bf r})\right]G\left({\bf r},{\bf r'};[\phi];N\right) = \delta
({\bf r}-{\bf r'})\delta(N) \quad.
\end{eqnarray}
It is convenient to make the Laplace transformation 
\begin{eqnarray}
G\left({\bf r},{\bf r'};[\phi];\mu\right) = \int\limits_0^\infty dN
G\left({\bf r},{\bf r'};[\phi];N\right)\exp(-\mu N) ,
\end{eqnarray}
after which the equation for $ G\left({\bf r},{\bf
    r'};[\phi];\mu\right)$ yields
\begin{eqnarray}
G\left({\bf r},{\bf r'};[\phi];\mu\right)
= \frac{\int D\psi \psi({\bf r})\psi({\bf r'})\exp \left\{-\frac{1}{2}\int
  d^dr\psi({\bf r})\left[\mu - \frac{a^2}{2 d}\nabla^2 + i\phi
  ({\bf r})\right]\psi({\bf r'})\right\}}{\int D\psi exp \left\{-\frac{1}{2}\int
  d^dr\psi({\bf r})\left[\mu - \frac{a^2}{2 d}\nabla^2 + i\phi
  ({\bf r})\right]\psi({\bf r'})\right\}}.
\label{GaussIntegr}
\end{eqnarray}
In order to avoid the denominator in eq.(\ref{GaussIntegr}) we should
upgrade the field $\psi$ by introducing $n$-component field
$\vec\psi=\{\psi_1,\psi_2,\dots \psi_n\}$. Then by using the replica
trick, we have  
\begin{eqnarray}
G\left({\bf r},{\bf r'};[\phi];\mu\right)=\lim_{n \to 0}\int
\prod_{a=1}^n D\psi_{\alpha}\psi_1({\bf r})\psi_1({\bf r'})\exp\left\{-\frac{1}{2}\sum_{a=1}^{n}\int
  d^dr\psi_{a}({\bf r})\left[\mu - \frac{a^2}{2 d}\nabla^2 + i\phi
  ({\bf r})\right]\psi_{a}({\bf r'})\right\}
\label{Correlator1}
\end{eqnarray}
One can make now the Laplace transformation from  both sides of
eq.(\ref{Partition1}) and substitute in it eq.(\ref{Correlator1}). The 
integration first over the field $\phi({\bf r})$ and then over
$\rho({\bf r})$ results in the following expression for grand canonical 
partition function of a polymer chain with ends fixed at points ${\bf
  r}$ and ${\bf r'}$:
\begin{eqnarray}
\Xi\left({\bf r},{\bf r'};\mu\right)=\lim_{n \to 0}\int
\prod_{a=1}^n D\psi_{a}\psi_1({\bf r})\psi_1({\bf
  r'})\exp\left\{ - H_n \left[\vec\psi ; \mu\right]\right\} ,
\end{eqnarray}
where
\begin{eqnarray}
H_n \left[\vec\psi ; \mu\right] = \frac{1}{2}\int
  d^dr \sum_{a=1}^{n}\psi_a ({\bf r})\left[\mu - \frac{a^2}{2 d}\nabla^2 \right]\psi_a ({\bf r'}) +
\sum_{m=1}^{\infty}\frac{v_{m+1}}{(m+1)!}\int
d^dr \sum_{a=1}^n \left[\frac{1}{2}\psi_{a}^2({\bf r})\right]^{m+1} \quad. 
\end{eqnarray}

\section{The connection to a random model}

Here we prove that in a poor solvent the polymer problem (see
eqs.(\ref{corr1})  and (\ref{hamilt})) can be connected to {\it a
one-component random} model. We shell restrict our consideration only
to second and third virial coefficients. The generalization to the
case with an arbitrary number of virial coefficients is
straightforward.

Let us consider the one component random model with Hamiltonian
\begin{eqnarray}
{\cal H} \{\psi\} = \frac{1}{2}\int\, d^{d}r\left[ \mu \psi^{2}(r) +
  \frac{a^2}{2d} \left(\nabla\psi\right)^2 + t(r) \psi^{2} (r) \right] 
, \label{hamiltAp}
\end{eqnarray}
where $t(r)$ is a non-Gaussian random field with the generating functional
\begin{eqnarray}
\Phi \{\rho (r)\} &\equiv& \int Dt(r) P\{t(r)\}\exp\left\{-\int
  d^drt(r)\rho (r)\right\}\nonumber\\
&=& \exp\left\{ \frac{|v|}{8}\int\, d^{d}r\, \rho^{2}(r) -
  \frac{w}{3!8}\int\, d^{d}r \, \rho^{3}(r)\right\}.\label{GFAp}
\end{eqnarray}
In eq.(\ref{GFAp}) $P\{t(r)\}$ is the distribution functional of the
field $t(r)$. One can easily check that the replication of the
Hamiltonian (\ref{hamiltAp}) and the subsequent averaging over $t(r)$, 
i.e.
\begin{eqnarray}
\overline{Z^n} = \overline{\int \prod_{a=1}^n D\psi_a \exp
  \left\{-\sum_{a=1}^n {\cal H}\{\psi_a \}\right\}},
\label{partfunkAP}
\end{eqnarray}
leads to the effective replicated Hamiltonian (\ref{hamilt}).

>From the probabilistic interpretation (\ref{GFAp}) one can explicitly
find the central moments of $t(r)$. The expansion of both sides of
eq.(\ref{GFAp}) yields
\begin{eqnarray}
&&\sum_{m=0}^{\infty}\frac{(-1)^m}{m!}\int dr_1\dots dr_m
\overline{t(r_1)t(r_2)\dots
  t(r_m)}\rho(r_1)\rho(r_2)\dots\rho(r_m)\nonumber\\
&=&\sum_{k=0}^{\infty}\frac{1}{k!}\sum_{l=0}^k
\frac{k!(-1)^l}{l!(k-l)!}\left[\frac{|v|}{8}\int d^dr
  \rho^2(r)\right]^{k-l}\left[\frac{w}{3!8}\int d^dr
  \rho^3(r)\right]^l \quad.
\label{expansionAp}
\end{eqnarray}
By making the $m$-terms of $\rho (r)$ on both sides of
eq.(\ref{expansionAp}) equal, one gets
\begin{eqnarray}
&&\int dr_1\dots dr_m
\overline{t(r_1)t(r_2)\dots
  t(r_m)}\rho(r_1)\rho(r_2)\dots\rho(r_m)\nonumber\\
&=&\sum_{k=\{m/3\}}^{[m/2]}\frac{m!}{(m-2k)!(3k-m)!}
\left[\frac{|v|}{8}\int d^dr
  \rho^2(r)\right]^{3k-m}\left[\frac{w}{3!8}\int d^dr
  \rho^3(r)\right]^{m-2k},
\label{expansion1Ap}
\end{eqnarray}
where $[m/2]$ stands for the greatest integer number less then $m/2$
and \{m/3\} is the smallest integer number larger then $m/3$. By
making use the representations $\int d^dr \rho^2(r)=\int
d^dr_1d^dr_2\delta({\bf r}_1-{\bf r}_2)\rho({\bf r}_1)\rho({\bf r}_2)$
and $\int d^dr \rho^3({\bf r})=\int
d^dr_3d^dr_4d^dr_5\delta({\bf r}_5-{\bf r}_4)\delta({\bf r}_5-{\bf
  r}_3)\rho({\bf r}_3)\rho({\bf r}_4)\rho({\bf r}_5)$ 
one finally gets:
\begin{eqnarray}
&&\overline{t(r_1)t(r_2)\dots t(r_m)}=
\sum_{k=\{m/3\}}^{[m/2]}\frac{m!}{(m-2k)!(3k-m)!}\left(\frac{|v|}{8}\right)^{3k-m}\left(\frac{w}{3!8}\right)^{m-2k}\nonumber\\
&\times& \delta({\bf r}_1-{\bf r}_2)\delta({\bf r}_3-{\bf r}_4)\dots
\delta({\bf r}_{6k-2m-1}-{\bf r}_{6k-2m})\nonumber\\
&\times&\delta({\bf r}_{6k-2m+1}-{\bf r}_{6k-2m+2})\delta({\bf
  r}_{6k-2m+1}-{\bf r}_{6k-2m+3})\dots \delta({\bf r}_{m-2}-{\bf
  r}_{m-1})\delta({\bf r}_{m-2}-{\bf r}_{m}).
\label{momentsAp}
\end{eqnarray}
In eq.(\ref{momentsAp}) the second line includes $3k-m$
$\delta$-functions with arguments successively pairwise divided
between $6k-2m$ points. The third line includes $m-2k$
$\delta$-functions so that in each successive pairs of them one
$\bf{r}$ - point is common. 

Let us consider some particular cases of eq.(\ref{momentsAp}).

i)$m=1$. Then $\{m/3\}=1$, $[m/2]=0$ and
\begin{eqnarray}  
\overline{t(r)} =0
\label{moment1Ap}
\end{eqnarray}
ii)$m=2$. Then $\{m/3\}=1$, $[m/2]=1$ and eq.(\ref{momentsAp}) reads
\begin{eqnarray}  
\overline{t({\bf r}_1)t({\bf r}_2)} =\frac{|v|}{4}\delta({\bf r}_1-{\bf 
r}_2)
\label{moment2Ap}
\end{eqnarray}
iii) At $m=3,\{m/3\}=1$, $[m/2]=1$ (i.e. k=1) and one gets 
\begin{eqnarray}  
\overline{t({\bf r}_1)t({\bf r}_2)t({\bf r}_3)} =\frac{w}{8}\delta({\bf 
r}_1-{\bf r}_2)\delta({\bf r}_1-{\bf r}_3)
\label{moment3Ap}
\end{eqnarray}
iv) At $m=4,\{m/3\}=2$, $[m/2]=2$ (i.e. k=2) and  
\begin{eqnarray}  
\overline{t({\bf r}_1)t({\bf r}_2)t({\bf r}_3)t({\bf r}_4)}
=\frac{3}{16}|v|^2\delta({\bf r}_1-{\bf r}_2)\delta({\bf r}_3-{\bf r}_4)
\label{moment4Ap}
\end{eqnarray}
v) Finally at $m=5,\{m/3\}=2, [m/2]=2$ and
\begin{eqnarray}  
\overline{t({\bf r}_1)t({\bf r}_2)t({\bf r}_3)t({\bf r}_4)t({\bf
      r}_5)} =\frac{5!}{3!64}|v|w\delta({\bf r}_1-{\bf r}_2)\delta({\bf 
  r}_3-{\bf r}_4)\delta({\bf r}_3-{\bf r}_5) \quad.
\label{moment5Ap}
\end{eqnarray}  
The important feature of these moments is that all of them are
positive, which  means that $t(r)$ is real.

\section{Spatial fluctuations for weakly inhomogeneous globule}
In this appendix we give the Landau-expansion only for the case when
$q$ and coefficients $\Gamma^{(2)} , \Gamma^{(3)} , \Gamma^{(4)}$ are
weakly ${\bf k}$ - dependent in the RS - sector. The spatial Fourier
transformation in eq.(\ref{LandauExp}) leads to the following
effective Hamiltonian
\begin{eqnarray}
H_{\rm RS}\{q({\bf k})\} &=& \int \frac{d^d k}{(2\pi)^d}\Gamma^{(2)}
({\bf k})q({\bf k})q({\bf- k}) +  \int \frac{d^d k_1d^d k_2}{(2\pi)^2d}\Gamma^{(3)}
({\bf k_1},{\bf k_2})q({\bf k_1})q({\bf k_2})q({\bf -k_1 -
  k_2})\nonumber\\
&+& \int \frac{d^d k_1d^d k_2 d^d k_3}{(2\pi)^3d}\Gamma^{(4)}
({\bf k_1},{\bf k_2},{\bf k_3})q({\bf k_1})q({\bf k_2})q({\bf k_3})q({\bf -k_1 -
  k_2} -  {\bf k_3})  + \dots   \quad,
\label{ExpansionApp}
\end{eqnarray}
where 
\begin{eqnarray}
\Gamma^{(2)}({\bf k}) = \frac{2}{|v|} - \int \frac{d^d
  \kappa}{(2\pi)^d} \;\frac{1}{\left[\frac{a^2}{2d}\kappa^2 + \mu
  \right]\left[\frac{a^2}{2d}({\bf \kappa} - {\bf k})^2 + \mu \right]}\quad,
\label{Gamma2}
\end{eqnarray}
\begin{eqnarray}
\Gamma^{(3)}({\bf k_1},{\bf k_2}) = \frac{4}{3}\left[ \frac{w}{|v|^3} 
  - \int \frac{d^d
  \kappa}{(2\pi)^d} \;\frac{1}{\left[\frac{a^2}{2d}\kappa^2 + \mu
  \right]\left[\frac{a^2}{2d}({\bf \kappa} - {\bf k}_1)^2 + \mu
  \right]\left[\frac{a^2}{2d}({\bf \kappa} - {\bf k}_1 - {\bf k}_2)^2 + \mu
  \right]   }\right] \quad,
\label{Gamma3}
\end{eqnarray}
\begin{eqnarray}
&&\Gamma^{(4)}({\bf k_1},{\bf k_2},{\bf k_3}) =
\frac{2}{|v|^4}\left(\frac{w^2}{|v|} + \frac{z}{3}\right) 
\nonumber\\
&&-2\int \frac{d^d
  \kappa}{(2\pi)^d} \;\frac{1}{\left[\frac{a^2}{2d}\kappa^2 + \mu
  \right]\left[\frac{a^2}{2d}({\bf \kappa} - {\bf k}_1)^2 + \mu
  \right]\left[\frac{a^2}{2d}({\bf \kappa} - {\bf k}_1 - {\bf k}_2)^2 + \mu
  \right]\left[\frac{a^2}{2d}({\bf \kappa} - {\bf k}_1 - {\bf k}_2 - {\bf k}_3)^2 + \mu
  \right]   }\quad.
\label{Gamma4}
\end{eqnarray}
For weak fluctuations around MF - solution $q_m$ to be investigated
one should estimate Hessian
\begin{eqnarray}
\left.\frac{\delta^2 H_{\rm RS}}{\delta q({\bf r}_1)\delta q({\bf
      r}_2)}\;\right|_{q = q_{m}} = 2\Gamma^{(2)}({\bf r}_1 - {\bf
  r}_2)  + 6 q_m \int d^d  r_3 \Gamma^{(3)}({\bf
  r}_2 -  {\bf r}_1, {\bf
  r}_3 -  {\bf r}_1)\nonumber\\
+ 12 q_m^2 \int d^d  r_3 d^d  r_4 \Gamma^{(4)}({\bf
  r}_2 -  {\bf r}_1, {\bf
  r}_3 -  {\bf r}_1, {\bf
  r}_4 -  {\bf r}_1) + \dots \quad,
\label{Hessian1}
\end{eqnarray}
or in the Fourier space
\begin{eqnarray}
\left.\frac{\delta^2 H_{\rm RS}}{\delta q({\bf r}_1)\delta q({\bf
      r}_2)}\;\right|_{q = q_{m}} = 2\Gamma^{(2)}({\bf k}) + 6 q_m
\Gamma^{(3)}({\bf k}, {\bf k}=0) + 12 q_m^2\Gamma^{(4)}({\bf k},  {\bf
  k}=0,  {\bf k}=0) + \dots  \quad.
\label{Hessian2}
\end{eqnarray}
The resulting effective Hamiltonian  expansion around the MF - solution takes the following form
\begin{eqnarray}
H_{\rm RS}\{q_m, \Delta q({\bf k})\} &=& H_{\rm RS}\{q_m\} + \frac{1}{2} 
\;\int \left.\frac{d^d k}{(2\pi)^d}\; \frac{\delta^2 H_{\rm RS}}{\delta
  q({\bf k})\delta q({\bf -k})}\right|_{q = q_m}\; \Delta q({\bf
k})\Delta q({\bf -k})\nonumber\\
&=& A q_m^2 + Bq_m^3 + C q_m^4 \nonumber\\
&+&\int\frac{d^d k}{(2\pi)^d}\left[\Gamma^{(2)}({\bf k}) + 3 q_m
\Gamma^{(3)}({\bf k}, {\bf k}=0) + 6 q_m^2\Gamma^{(4)}({\bf k},  {\bf
  k}=0,  {\bf k}=0)\right]\nonumber\\
&&\times\Delta q({\bf
k})\Delta q({\bf -k}) \quad,
\label{Fluctuations}
\end{eqnarray}
where 
\begin{eqnarray}
\Delta q({\bf k}) = q({\bf k}) - q_m
\end{eqnarray}
and
\begin{eqnarray}
A = \Gamma^{(2)}(0) \quad,\quad B = \Gamma^{(3)}(0 , 0)\quad,\quad C = 
\Gamma^{(4)}(0 , 0 , 0) \quad.
\label{ABC_Abstr}
\end{eqnarray}
The expressions for A, B and C are given by eqs. (\ref{coeff1})
-(\ref{coeff3}). It is easy to estimate the integrand in
eq.(\ref{Fluctuations}) at small ${\bf k}$ (weak inhomogeneity). The
straightforward calculations yields
\begin{eqnarray} 
H_{\rm RS}\{q_m, \Delta q({\bf k})\} = H_{\rm RS}\{q_m\} + \int
\frac{d^d k}{(2\pi)^d}\;\{X(|v|,w) + (ka)^2 Y(|v|,w)\}\Delta q({\bf
k})\Delta q({\bf -k}),
\label{Fluctuations1}
\end{eqnarray}
where
\begin{eqnarray}
X = A + 3 q_m B + 6 q_m^2 C
\label{X}
\end{eqnarray}
and
\begin{eqnarray}
Y &=& \left(\frac{1}{2\pi d a^2}\right)^{d/2} \; \frac{2d -
  3}{12\mu^{\frac{6 -  d}{2}}}\; \Gamma\left(\frac{6 - d}{2}\right) +
q_m \left(\frac{1}{2\pi d a^2}\right)^{d/2} \; \frac{3d -
  4}{9\mu^{\frac{8 -  d}{2}}}\; \Gamma\left(\frac{8  -
    d}{2}\right)\nonumber\\
&+& q_m^2 \left(\frac{1}{2\pi d a^2}\right)^{d/2} \; \frac{4d -
  5}{20\mu^{\frac{10 -  d}{2}}}\; \Gamma\left(\frac{10  -
    d}{2}\right) \quad.
\label{Y}
\end{eqnarray}
>From (\ref{Fluctuations1}) it is obvious that the corresponding
correlation function 
\begin{eqnarray}
\left<\left|\Delta q({\bf k})\right|^2\right> = \frac{1}{X(|v|,w) +
  (ka)^2 Y(|v|,w)} \quad.
\label{Correlator_q}
\end{eqnarray}

\end{appendix}

\end{document}